\documentclass[prb,singlecolumn,a4paper,amsmath,amssymb]{revtex4}

\usepackage{geometry}
\usepackage{amsfonts}
\usepackage{latexsym}
\usepackage{graphicx}
\usepackage[usenames]{color}
\usepackage[utf8]{inputenc}
\usepackage{hyperref}
\usepackage{caption}
\usepackage{morefloats}
\usepackage{epstopdf}
\usepackage{systeme}
\usepackage{listings}
\usepackage{array}
\usepackage{circuitikz}
\usepackage{tikz}

\newcommand{\be}{\begin{equation}}
\newcommand{\ee}{\end{equation}}

\newcommand{\kb}{k_\mathrm{_B}}

\begin{document}

\title{The Day of the Inductance. Problem of the 7$^\mathrm{th}$ Experimental Physics Olympiad, Skopje, 7 December 2019}


\author{Todor~M.~Mishonov}
\email[E-mail: ]{mishonov@bgphysics.eu}
\affiliation{Physics Faculty, St.~Clement of Ohrid University at Sofia, 
5 James Bourchier Blvd., BG-1164 Sofia}
\affiliation{Institute of Solid State Physics, Bulgarian Academy of Sciences
72 Tzarigradsko Chaussee Blvd., BG-1784 Sofia, Bulgaria}
\author{Riste Popeski-Dimovski}
\email{ristepd@gmail.com}
\author{Leonora Velkoska}
\affiliation{Institute of Physics, Faculty of Natural Sciences and Mathematics, 
``Ss. Cyril and Methodius'' University, Skopje, R. Macedonia}
\author{Iglika~M.~Dimitrova}
\affiliation{Faculty of Chemical Technologies, Department of Physical Chemistry,
University of Chemical Technology and Metallurgy,
8 Kliment Ohridski Blvd., BG-1756 Sofia}
\author{Vassil~N.~Gourev}
 \affiliation{Department of Atomic Physics, Physics Faculty, St.~Clement of Ohrid University at Sofia, 
5 James Bourchier Blvd., BG-1164 Sofia}
\author{Aleksander~P.~Petkov, Emil~G.~Petkov, Albert~M.~Varonov}
\affiliation{Institute of Solid State Physics, Bulgarian Academy of Sciences
72 Tzarigradsko Chaussee Blvd., BG-1784 Sofia, Bulgaria}%

\date{December 28, 2019}

\begin{abstract}
This is the problem of the 7$^\mathrm{th}$ international Experimental Physics Olympiad ``The Day of the Inductance''.
A large inductance made by a general impedance converter is measured by the Maxwell-Wien bridge.
The youngest students (S category) have the problem to measure the parameters of the elements in the experimental set-up.
The students at middle age (category M) should balance a Wheatstone bridge with resistors only.
The inductance measurement is supposed to be accomplished by the students from the upper age (L category) only.
From the university students (XL category) we expect error estimation and professionally made measurement protocol.
As a homework problem (Sommerfeld price) the general formula for the impedance of the general impedance converter, where the general equation for operational amplifiers is included, has to be derived.
Problems for further work are also given in the text.
The described set-up can be considered as a prototype for companies producing laboratory equipment.
\end{abstract}
\captionsetup{labelfont={normalsize},textfont={small},justification=centerlast}
\maketitle

\section{Introduction}

From its very beginning, the Experimental Physics Olympiad (EPO) is worldwide known;
all Olympiad problems have been published in Internet~\cite{EPO1,EPO2,EPO3,EPO4,EPO5,EPO6} and from the very beginning there were 120 participants.
In the last years high-school students from 10 countries participated 
and the distance between the most distant cities is more than 4~Mm.

Let us describe the main differences between EPO and other similar competitions.
\begin{itemize}
\item Each participant in EPO receives as a gift from the organizers the set-up, 
which one worked with.
So, after the Olympiad has finished, even bad performed participant is able to repeat the experiment and reach the level of the champion.
In this way, the Olympiad directly affects the teaching level in the whole world.
After the end of the school year, the set-up remains in the school, where the participant has studied.
\item Each of the problems is original and is connected to fundamental physics or the  understanding of the operation of a technical patent.
\item The Olympic idea is realized in EPO in its initial from 
and everyone willing to participate from around the world can do that.
There is no limit in the participants number.
On the other hand, the similarity with other Olympiads is that the problems are direct illustration of the study material and alongside with other similar competitions mitigates the secondary education degradation, which is a world tendency.
\item One and the same experimental set-up is given to all participants but the tasks are different for the different age groups, the same as the swimming pool water is equally wet for all age groups in a swimming competition.
\item One of the most important goals of the Olympiad is the student to repeat the experiment at home and to analyze the theory necessary for the understanding.
In this way any even badly performed motivated participant has the possibility to be introduced to the corresponding physics field, even though there is no physics classroom in his/her school, even though the physics education in his/her country to be deliberately destroyed.
\end{itemize}

We will briefly mention the problems of former 6 EPOs: 
1) The setup of EPO1 was actually a student version of the American patent for auto-zero
and chopper stabilized direct current amplifiers.~\cite{EPO1}
It was notable that many students were able to understand the operation of an American patent without special preparation.\cite{chopper}
2) The problem of EPO2~\cite{EPO2} was to measure Planck constant by diffraction of a LED light by a compact disk.
3) A contemporary realization of the assigned to NASA patent for the use of negative impedance converter for generation of voltage oscillations was the set-up of EPO3.~\cite{EPO3}
4) EPO4~\cite{EPO4} was devoted to the fundamental physics -- to determine the speed of light by measuring electric and magnetic forces.
The innovative element was the application of the catastrophe theory in the analysis of the stability of a pendulum.
5) The topic of the EPO5~\cite{EPO5} was to measure the Boltzmann constant $\kb$ following the Einstein idea of study thermal fluctuations of electric voltage of a capacitor.
6) The EPO6~\cite{EPO6} problem can be considered as a continuation of the previous Olympiad.
With a similar electronic circuit Schottky noise is measured and his idea for the determination of the electron charge is realized.
Each problem given at a EPO can be considered as a dissertation in methodology of physics education.

In short, the established traditions is a balance between contemporary working technical inventions and fundamental physics.

The EPO problems are meant for high school and university students but are posed by teachers with co authorship with  colleagues working in universities or scientific institutes.
For colleagues interested in new author's problems for the needs of the contemporary physics education we share our experience in the description of the experimental set-ups described at a university level.
These are for instance:
1) The determination of the Planck constant without light but only with electronic processes study;~\cite{EJP_Planck} this set-up requires the usage of an oscilloscope but in some countries the oscilloscopes are available in  in the high schools physics labs and the prices of the former is constantly going down.
The speed of the light without the usage of scales or high frequency equipment is another innovative set-up~\cite{EJP_light} for high school education.
And the idea for this experiment is given by our teacher in electrodynamics Maxwell.

In the physics curriculum in all countries it is mentioned that the temperature is a measure of average kinetic energy of the gas molecules but the Boltzmann constant $k_\mathrm{B}$ that gives the relation between energy and temperature is not measured in high school and even rarely in the best universities.
The experimental set-up for $k_\mathrm{B}$ by the method proposed by Einstein (the EPO5 problem)
is described~\cite{EJP_Boltzmann} as a set-up for university school lab exercise in a impact factor journal.
But what larger impact an experimental set-up that is used by high school students from Kazakhstan and Macedonia and the surrounding countries can have.
More than 100 set-ups were distributed around the world.

Similar thoughts can be expressed for the electron charge $q_e$.
This fundamental constant is also mentioned in the high school education as a humanitarian incantation but is not measured.
We broke this tradition and described an experimental set-up (from EPO6) in the European Journal of Physics.~\cite{EJP_Schottky}
This set-up can be built for a week in every high school.
The Schottky idea for determination of the electron charge by measuring voltage fluctuations is used.
From the idea to the realization more than 100 years have passed and one of the reasons is that in many countries the largest enemy of the education is the ministry of education.
The mission of the physics teachers in the worldwide progress is evident -- precisely our science reshaped the world in the last century.
Successful innovative EPO set-ups after some update can be manufactured by companies specialized in production of educational equipment like TeachSpin\cite{TeachSpin} for instance. 

The translation of present text in Bulgarian, Macedonian, Russian and Serbian can be found in the auxiliary folder of the present arXiv; corresponding tex files have to be download and compiled together with files for figures.
In the same folder can be fond the scanned work
of the absolute champion Dimitar Jorlev.

\section{Olympiad Experimental Problem}

\begin{figure}[h]
\centering
\includegraphics[scale=0.4]{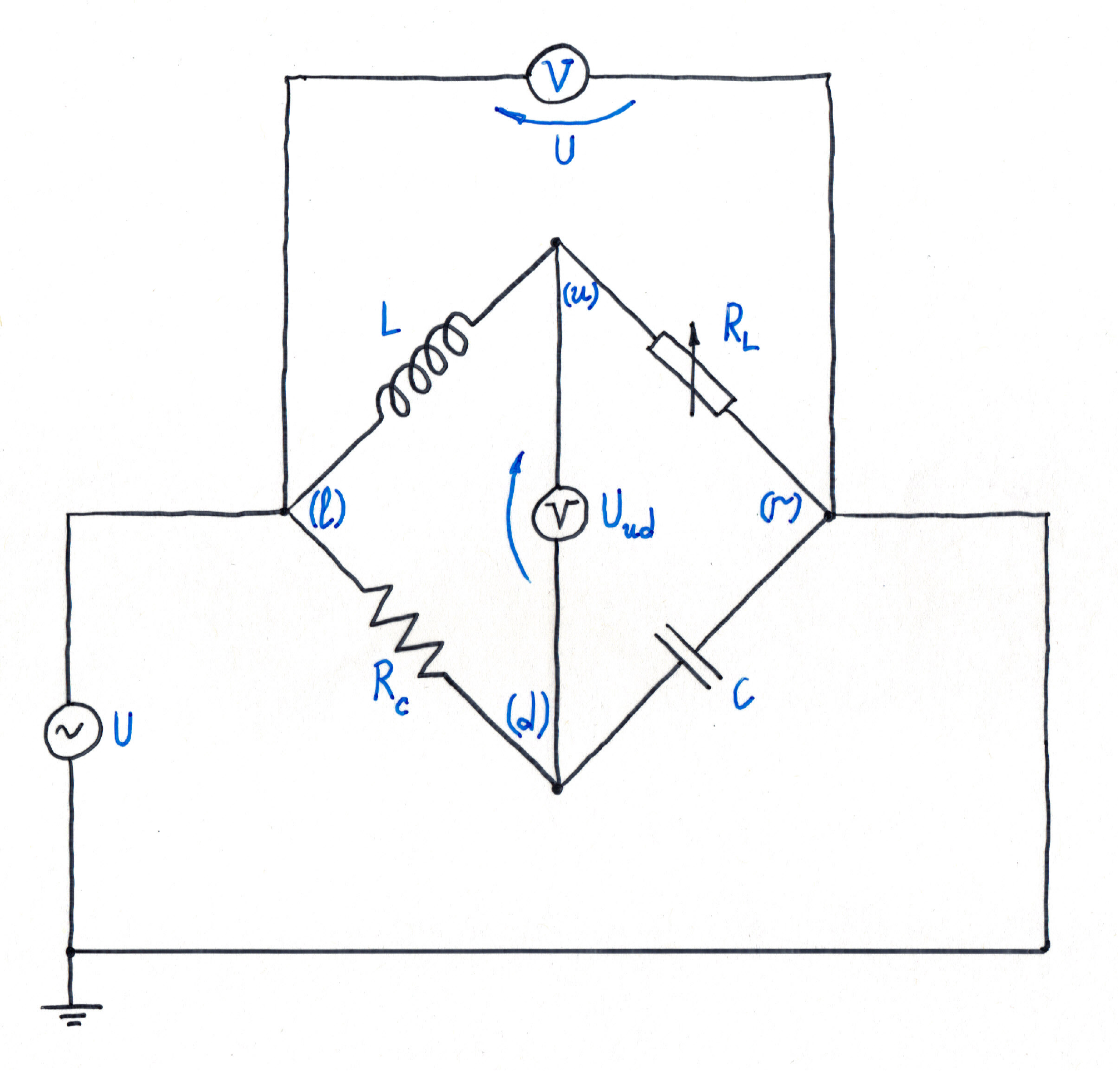}
\caption{Maxwell bridge for inductance measurement $L$ at known capacitor capacity $C$, resistor resistance $R_C$ and measured potentiometer resistance $R_L$.
The external voltage $U$ with frequency $f=\omega/2\pi$ is applied between the pins $(l)$ and $(r)$.
By rotating the potentiometer axis the voltage $U_\mathrm{ud}$ between the pins (u) and (d) is minimized.
To improve the bridge balance, switch the voltmeter to maximum precision.
At the end measure the potentiometer resistance and use Eq.~(\ref{inductance}).
When disconnecting the potentiometer, be careful not to touch its axis.}
\label{Fig:set-up}
\end{figure}

The circuit of the experimental set-up for measurement of inductance by a bridge described by Maxwell is shown in Fig.~\ref{Fig:set-up} and a photograph of the set-up you have to put together to measure the given in the set inductance by the equation
\begin{equation}
\label{inductance}
L= C R_C R_L.
\end{equation}
\begin{figure}[h]
\centering
\includegraphics[scale=0.2]{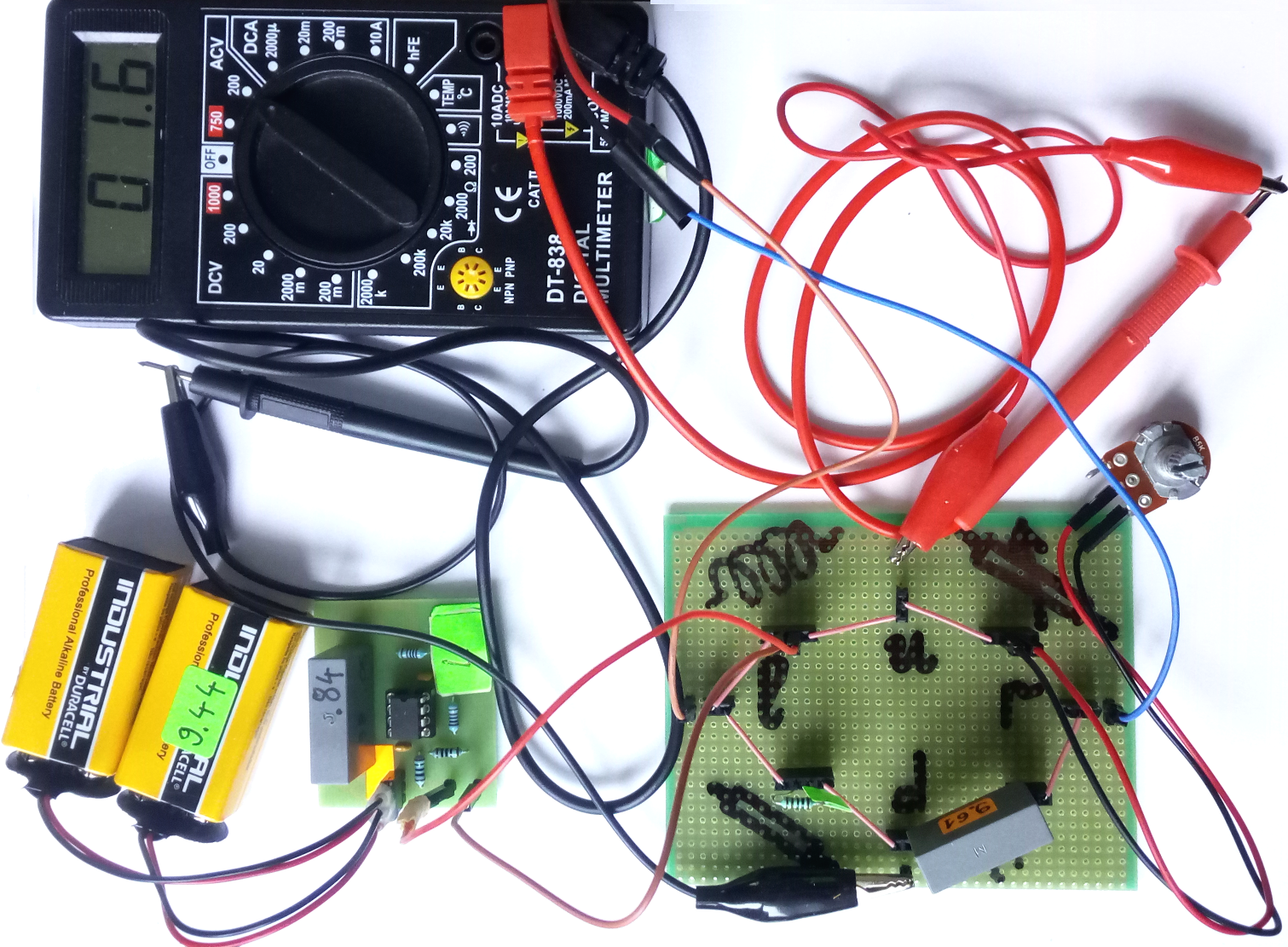}
\caption{A photograph of the experimental set-up presented schematically in Fig.~\ref{Fig:set-up}.
In the initial problem for category (M) the inductance (up left) is replaced by a resistor $R_1$, and the capacitor (down right) with $R_3$.
The given resistor set is in the small plastic bag.
Let the resistor down left to be with the largest resistance $R_4$ initially.
Confer Fig.~\ref{Fig:bridge}
where $R_1=Z_1,$ $R_3=Z_3$ and $R_4=Z_4$
capital letters denote place in the bridge schematics.
The wires of the external voltage applied to the bridge horizontally between the pins (l) and (r) are also seen.
The balance of the bridge is accomplished by minimizing the vertical voltage between the pins (u) and (d).
}
\label{Fig:foto}
\end{figure}

The inductance itself is implemented by integral circuits, that are powered with 9~V batteries and is connected to the bridge with two other of the given cables, see Fig.~\ref{Fig:foto}.
At connecting the batteries you should be careful the orange label to be set against an orange label, at wrong polarity the used integral circuits (operational amplifiers) burn!

The resistance $R_C$ from the down side of the bridge is fixed.
You have to use resistances from the given set for the different measurements.
For resistance $R_L$ from the upper side of the bridge use the given potentiometer.

Attach the capacitor $C$ with nominal capacity 10~$\mu$F, its more accurate measured value is written in $\mu$F, on the down right branch of the bridge.

For a source of voltage use the cables reaching your working place, at repeating the experiment at home use a transformer and a voltage divider so that the voltage to be less than 5~V at frequency $f \leq 60$~Hz.

If you have only an ordinary multimeter switch it to the smallest constant voltage (DC), 200~mV for instance.
If you have a precise voltmeter for alternating voltage (AC), the bridge balance i.e. the minimizing of the voltage $U_\mathrm{ud}$ between the up (u) and down (d) pins the bridge may be executed with  mV precision.

In short, at fixed $R_C$ you change $R_L$ to minimize $U_\mathrm{ud}$
or its fluctuations (random values around the average) if you use DC voltmeter and determine how many Henry (H) is the given inductance $L$.
The main criterion for the evaluation of your work at the Olympiad is the reliability of the determination of the inductance $L$.
If you have an idea how to accomplish this measurement work and describe the result from your measurement without reading the additional guiding text.
We do not need humanitarian text, we will examine your protocol for clearly made tables, graphs and result from the measurement.
The theoretical problems are given with the purpose to check the understanding of what is measured in cases of equal quality in experiment.

We expect a larger part of the participants to have modest practice in conducting measurements and experimental data processing.
That is why we offer a series of experiments, which are accessible for the youngest participants in category (S).

\section{Initial easy tasks. S}
\label{Sec:beginning}

\begin{enumerate}
\item Turn on the multimeter as a DCV and measure the voltages of the given 9~V batteries.
Such batteries are inside your multimeters, too.

\item Turn on the incoming to you alternating voltage to the left (l) and the right (r) pins of the bridge from the set-up.
Turn on the multimeter as an ACV voltmeter and measure this alternating voltage $U$ between these two pins, using the ``crocodile'' cables ``biting'' the remaining free pins.
This voltage has already been applied between the left (l) and the right (r) pins of the bridge, which you have to be balanced using a potentiometer.

\item Turn on the multimeter as an ohmmeter ($\Omega$), measure the range of the given potentiometer, i.e. what resistances you can have by rotating its axis, 
$R_\mathrm{pot}\in (R_\mathrm{min}, R_\mathrm{max})$.

\item With maximum accuracy measure the given resistors $r_i$, $i=1,\dots$ 
and write down the results in a table.

\item How many mV is the difference between $\frac12\,$V and $\frac13\,\mathrm{V}$?

\section{Tasks M}

\item Take 3 resistors from your set, for instance the smallest ones $r_1,$ $r_2$ and $r_3$.
Mount them on the given bridge by twisting their electrodes.
Schematically this is presented in Fig.~\ref{Fig:bridge}, as
$Z_1=r_1$ (up left, middle 2 holes), 
$Z_3=r_2$ (down right, connects two triple board connectors),
$Z_4=r_3$ (down left, middle 2 holes).
Let us repeat:
in the three quadruple board connectors the electrodes have to be connected in the 2 middle holes.
The outer holes should be connected with the $\Pi$-shaped jumper wires given in the set.

\item Follow in Fig.~{Fig:foto} how the 7 $\Pi$-shaped jumper wires should be mounted on the bridge.
Mount them and imagine that a closed circuit should be accomplished.

\item For resistor $Z_2 =R_\mathrm{pot}$ connect the given potentiometer using 2 of the given cables, see the photograph in Fig.~\ref{Fig:foto}.
Use the middle holes of the board connector on the upper right.
\begin{figure}[h]
\centering
\includegraphics[scale=1.0]{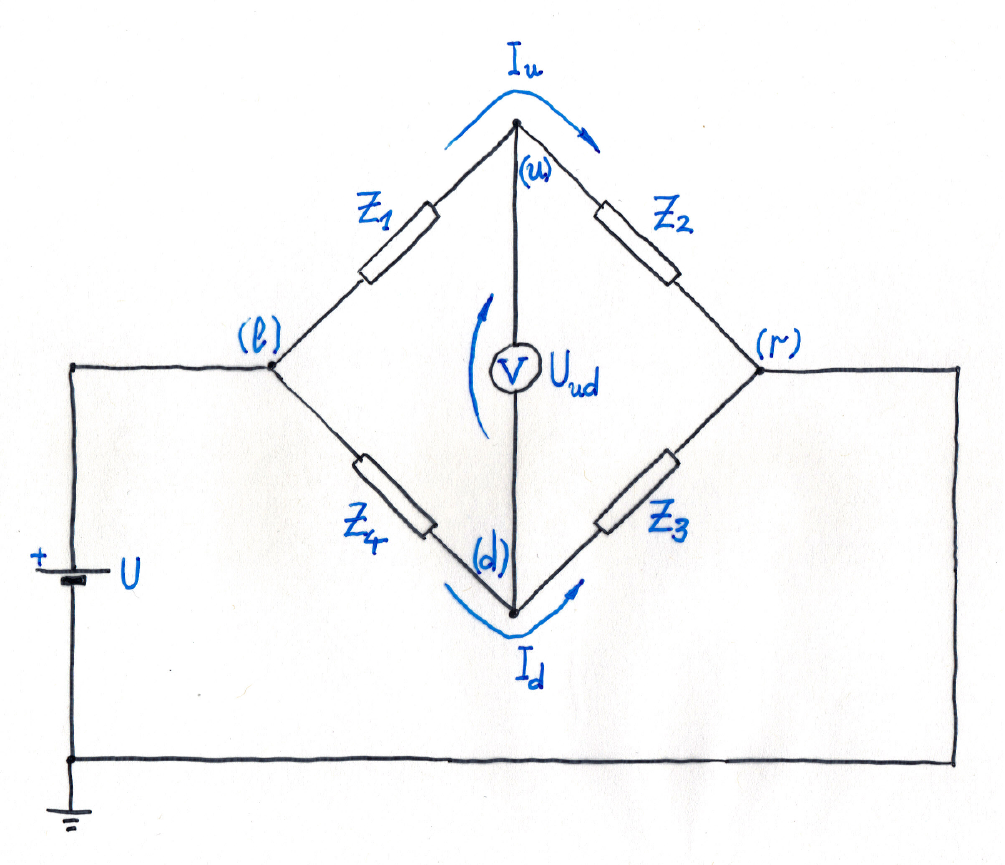}
\caption{Wheatstone bridge for resistance measurements by a bridge balance.
When the voltage $U_\mathrm{ud}$ is minimal by careful rotating the potentiometer $Z_2$ axis, we say that the bridge is balanced.
The resistors $Z_3$ and $Z_4$ are considered known.
Usually the resistor $Z_1$ is determined by the condition $Z_1=Z_2 Z_4/Z_3.$}
\label{Fig:bridge}
\end{figure}
\item Turn on a multimeter as a alternating-current voltmeter (ACV).
It should give the difference $U_\mathrm{ud}$ between the voltages in the up (u) and the down (d) pins of the bridge.

\item Turn on the external volatge between the left (l) and the right (r) pins of the bridge.
Turn on the potentiometer axis between the end positions and follow how $U_\mathrm{ud}$ changes. 
Now, when you have a general idea for the interval of changes of $U_\mathrm{ud}$, slowly rotate the potentiometer axis to minimize it.
If  $U_\mathrm{ud}\approx 0$ it is stated that the voltage $U_\mathrm{ud}$ is annulled and the bridge is balanced.

\item For this problem we do not have a sensitive galvanometer, but with a cheap multimeter only.
When the bridge is approximately balanced switch the multimeter as a direct current voltmeter (DCV) and try using its maximal sensitivity in the range of 200~mV.
When an alternating voltage is applied to the digital voltmeter for constant voltage, a random voltage from the interrupted sinusoid remains in the capacitors of the averaging circuit (inside the multimeter).
In this way, the digit on the display each second is different, but when the bridge is balanced the random values near the average (the fluctuations) are smaller.
Try improving the balance of the bridge by very carefully rotating the potentiometer axis aiming to minimize the fluctuations on the voltmeter display.
Follow the running digits and remember the minimal and maximal values of the voltage.
This difference should be minimized at different positions of the potentiometer axis.
In this way we simulate the operation of a sensitive and expensive galvanometer, which is rarely present operational in schools.

When you balance the bridge, be careful not to touch the potentiometer axis.
Disconnect the potentiometer and connect it to a multimeter in an ohmmeter $\Omega$ regime
with maximum allowed precision and write down the potentiometer resistance
$Z_2=R_\mathrm{pot},$ taken out from the up right position of the bridge.
The problem for bridge balancing with ohm resistances is central for the Olympiad because it illustrates an idea that can be realized in many other cases, for instance measurement of inductance which is the problem of the next category.

\item 
After you have measured the resistance of the potentiometer
$Z_2=R_\mathrm{pot}$
after you have balanced the bridge
and you have previously measured the resistances
placed in positions
$Z_1$, $Z_3$, and $Z_4$
calculate the ratios  
$Z_1/Z_2$, $Z_4/Z_3$,
their difference
$Z_1/Z_2-Z_4/Z_3$,
and most importantly the dimensionless number
\begin{equation}
\varepsilon=\frac{Z_1}{Z_2}-\frac{Z_4}{Z_3}.
\end{equation}
If you have time, replace the resistors,
balance the bridge and repeat the measurements again.
Use other resistors from the kit.
Represent the results in a table.

\section{Tasks L}

After you finish the category (S) and (M) problems connect the 9~V batteries to the inductance $L$ and measure it.

\item More precisely, connect the 9~V batteries to their holders (clips).

\item Now be careful, you can burn the integral circuit!
Orange label towards orange label; connect the batteries to the three pins of the inductance.

\item The inductance itself connect to the bridge from Fig.~\ref{Fig:foto} up left using the 2 middle holes of the 4-hole board connector.
Use the remaining 2 of the 4 given cables.

\item The capacitor connect down right, see again Fig.~\ref{Fig:foto}, as it is not necessary to apply much force.

\item Connect the potentiometer up right, as in the tasks for category (M).

\item Connect a resistor down left with resistance closest as possible to $1\,\mathrm{k}\Omega$.
Check whether there is anything forgotten and the bridge is a closed circuit.

\item Using the ``crocodiles'' connect a voltmeter between the pins 
(u) and (d). 
The multimeter initially has to be connected as an alternating current voltmeter ACV.

\item Rotate the potentiometer axis and try to minimize the voltage $U_\mathrm{ud}$.
This is a comparatively easy task because the multimeters in the alternating-current voltage (ACV) mode have little precision and the bridge is approximately balanced.

\item Now, the most complex part of the problem follows, do not lose patience and do not discourage.
These measurements may take an hour and require patience.
Switch the multimeter as a direct current voltmeter (DCV).
The suitable range you will find yourself.
The voltage changes constantly (fluctuates) and has random values in some interval $(U_\mathrm{min},\,U_\mathrm{max}).$

Carefully rotate the potentiometer axis, wait and see whether the difference $|U_\mathrm{max}-U_\mathrm{min}|$ has become less.
The goal of the measurement is to minimize this difference, for which optimism, patience and a little chance are required.

\item When you finish with the measurement, carefully disconnect the potentiometer being careful not to touch its axis and measure the resistance $R_\mathrm{pot}.$
After which the calculation of the inductance is a task for a calculator.
If you have time, repeat the measurements using the remaining resistors from the given set up to the end of the Olympiad and again calculate the inductance $L$ for each one of them.

Here the experimental (the most important) problems of the Olympiad end.
The theoretical problems follow, which give fewer points and are important at equal experimental achievements.

\section{Theoretical Problems. L}

\subsection{Bridge Balance}
\item A schematic of a bridge that is balanced in equilibrium, i.e. when a voltage $U$ is applied to its left (l) and right (r) pins, the voltage difference between its up (u) and down (d) pins $U_\mathrm{ud}=0$ is zero, is given in Fig.~\ref{Fig:bridge}.
Show that in these conditions we have equality between the relations of the resistances from the left and right side of the bridge
\begin{equation}
\label{bridge_balance}
r=\frac{Z_1}{Z_2}=\frac{Z_4}{Z_3}.
\end{equation}

\subsection{Maxwell Bridge Balance}
\item The Maxwell bridge for inductance measurement with alternating current is shown in Fig.~\ref{Fig:Maxwell}.
\begin{figure}[h]
\centering
\includegraphics[scale=0.3]{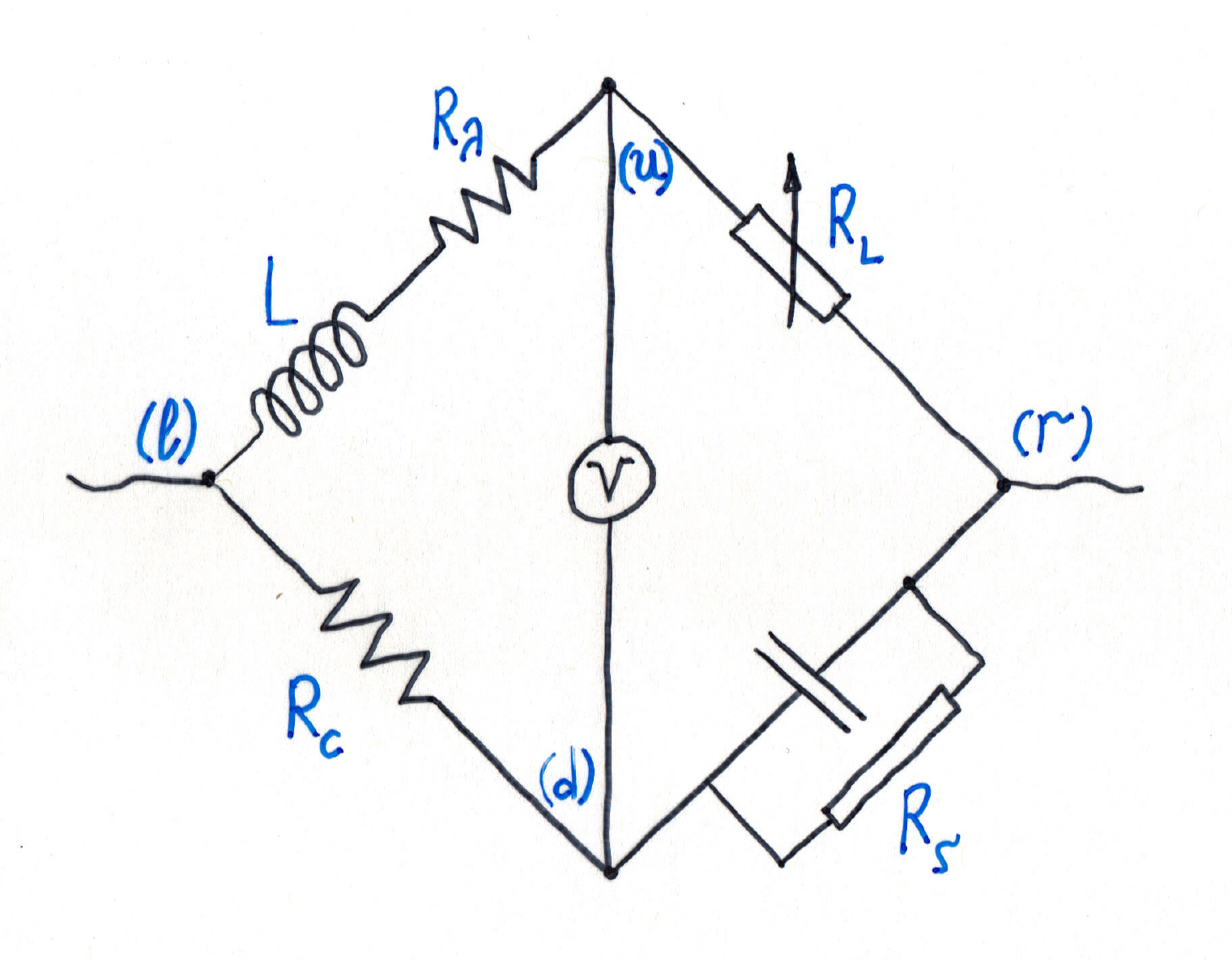}
\caption{Wien modification for the Maxwell bridge.
The voltage is applied between the pins (l) and (r) and when the voltage between the pins (u) and (d) is negligible, it is stated that the bridge is balanced and the balance conditions are $L/C=R_LR_C=R_\lambda R_\varsigma$.}
\label{Fig:Maxwell}
\end{figure}
When between the pins (l) and (r) an alternating (time dependent $t$) voltage 
\begin{equation} 
U(t)=U_0\sin(\omega t+ \varphi_0),
\end{equation}
where $U_0$ is the amplitude of the sinusoidal signal,
$\varphi_0$ is its initial phase and
$\omega=2\pi f$ is the angular frequency.
We can use the same condition for the bridge balance Eq.~(\ref{bridge_balance}), but applied for frequency dependent resistances $Z(\omega)$, which we call impedances.
For instance, for the circuit shown in  Fig.~\ref{Fig:Maxwell} we have real impedances
\begin{equation}
\label{z2_z4}
Z_2=R_L=R_\mathrm{pot},\qquad Z_4=R_C,
\end{equation}
confer Fig.~\ref{Fig:bridge}.
For our set-up $R_C$ is a metal layered resistor with fixed resistance and $R_L$ is a potentiometer whose middle electrode (leg) should be used.
For the other two branches of the bridge we should use the impedances of the inductances
$Z_1=Z_L$ and the capacity $Z_3=Z_C$, for which we have the general formulae
\begin{equation}
\label{zl_zc}
Z_L=\mathrm{j}\omega L,\qquad
Z_C=\frac1{\mathrm{j}\omega C}.
\end{equation}
In our treatment it is not necessary to use $\mathrm j=-1/\mathrm j$ and $\mathrm{j}^2=-1.$
Derive Eq.~(\ref{inductance}) for the inductance determination at known capacity $C$ and resistances $R_C$ and $R_L$ from the Maxwell bridge.
Illustrate the formulae with numerical example from a measurement of yours.

\subsection{General Impedance Converter, GIC}

\begin{figure}[h]
\centering
\includegraphics[scale=0.8]{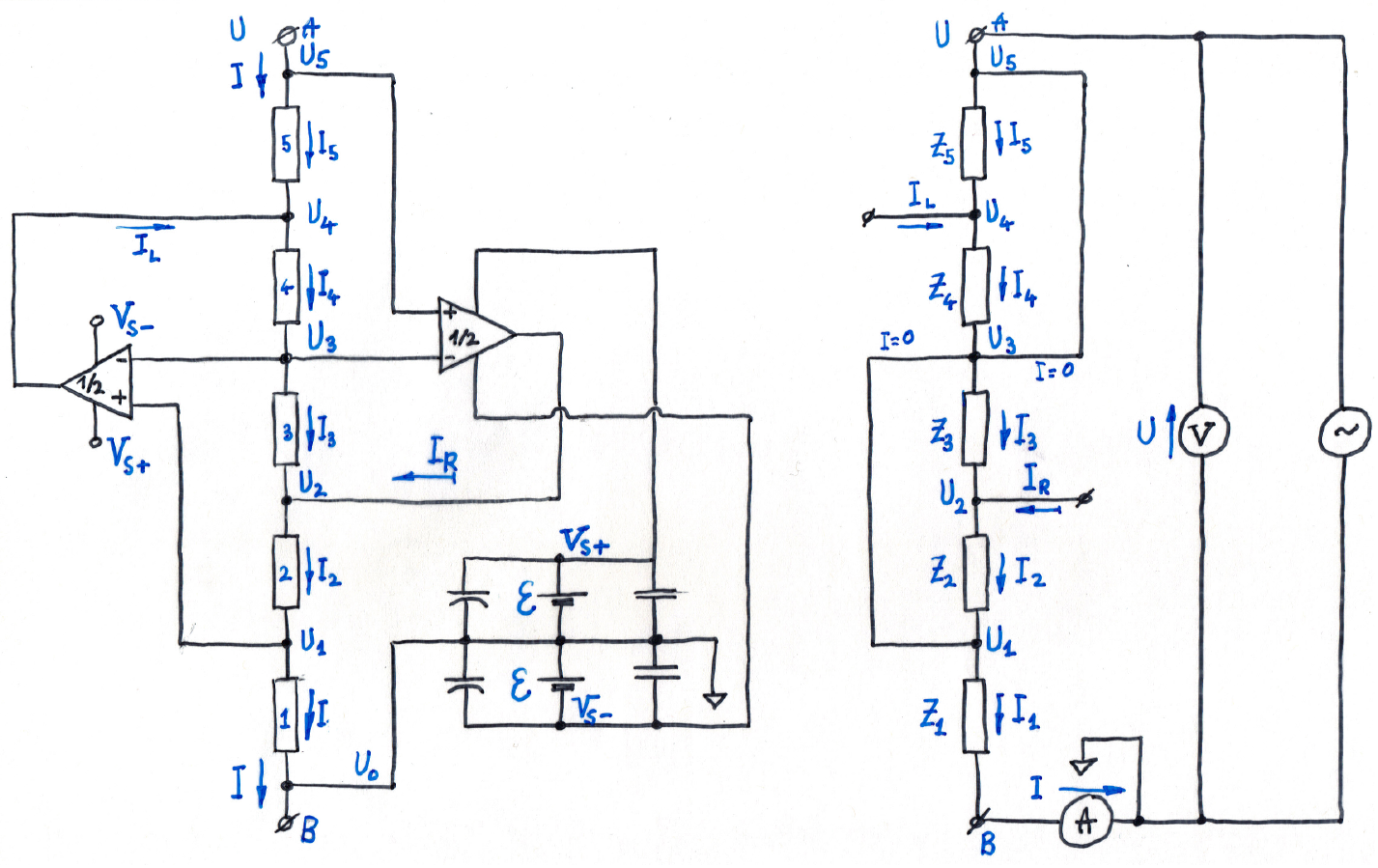}
\caption{General impedance converter, GIC.
Left: The GIC schematic used for the inductance from the experimental set-up of the problem.
The power supplies with 9~V batteries of the operational amplifiers and the capacitors parallely connected to the batteries are shown.
From the triangles vertices currents flow out so that the potentials of their voltage inputs $(+)$ and $(-)$ become equal.
Both operational amplifiers are made on an integral circuit and are packaged in a standard package in 8 electrodes.
Right: An effective circuit that alleviates the analysis.
The operational amplifiers and their power supply are removed.
The currents $I_L$ and $I_R$ flowing out from the operational amplifiers are presented as external.
Fictitious conductors between the points whose potentials are equaled by the operational amplifiers 
$U_5=U_3$ and $U_3=U_1$ are drawn.
It is shown that through these fictitious wires current does not flow. 
Ohm's law is applied for the impedances $Z_1,$ $Z_2,$ $Z_3,$ $Z_4$ and $Z_5$ and the formulae are the same, if these are resistors only.
The problem is the impedance to be calculated, i.e. the relation $Z=U/I$ between the voltage $U$ between the points (A) and (B) and the current, which flows between these electrodes.
The variables in this problem are independent from the same from the other sections.}
\label{Fig:GIC}
\end{figure}
\item The circuit of the general impedance converter is shown in Fig.~\ref{Fig:GIC}.
On the left schematic with triangles are denoted operational amplifiers that create such an exit current from the triangle vertices that the potentials denoted with (+) and (-) on their sides are equaled.
The batteries powering these operational amplifiers are also shown on the left figure.
On the right schematic these insignificant details are omitted.
The points with equaled potential are substituted with wires.
The power supply is also missing, the exit currents from the triangle vertices are represented as external currents incoming in the circuit.

The batteries, capacitors and the word operational amplifier are shown only to discourage the uncertain in themselves students.
For the same reason instead of a resistor, we use the word impedance, but the formulae are the same.
And the problem is much simpler than problems given in algebra homework.
When applying Ohm's law for the impedances $Z_1,$ $Z_2,$ $Z_3,$ $Z_4$ and $Z_5,$
find the relation $Z\equiv U/I$ between the input voltage$U$ and the input current $I$ shown in the schematic.

Instructions: Let us repeat the general rule for the exit current from the operational amplifiers:
from the triangle vertices such a current flows out that the potentials at the points (+) and (-) on their vertical sides are equaled.
For instance the current $I_L$ flowing out from the vertex of the left triangle equals the potentials in the points (3) and (5), i.e. $U_3=U_5.$
The currents flowing in the points (-) of both triangles are negligible and this leads to equality between the currents flowing through the impedances
$Z_3$ and $Z_4$, i.e. $I_3=I_4$.
Analogously, the zero current flowing in the (+) input of the left triangle according to Kirchoff's law for the charge conservation leads to the equality $I_1=I_2.$
For the potential at the point (B) it is convenient to choose $U_0=0$ (this point is grounded).
For the other input point of the circuit (A) for uniformity we choose the notions $U_5=U.$

Answer: 
\begin{equation}
\label{GIC_12345}
Z\equiv\frac{U}{I}=\frac{Z_1Z_3Z_5}{Z_2Z_4},
\end{equation}
do the derivation!

\item{How the general impedance converter can be used for creation of large inductances with very good quality, using only resistors and capacitors?}

\section{Homework Problem: General formula for the general impedance converter. XL}

\item General setup. 
At difficulty level this problem corresponds to student level but every high school student with practice in algebra can perform the derivation in a day.
The difficulty is that you are tired enough by the Olympiad.
Take a rest and try to continue working.
The solution should be sent during the night after the Olympiad until sunrise to the Olympiad email address epo@bgphysics.eu.
The best solution will win the Sommerfeld price with monetary equivalent of DM137.
The sum is given only personally on the morning at 11:00 (8 December 2019), but if you have left, the prize is co-authorship in the solution of the problem, which will be published in the space of a week.

Introduction: The equaling of the potentials at the input of the operational amplifier is only a low frequency approximation known as golden rule for the operational amplifiers.
In the general case we have the relation\cite{master} between the input and output voltages\begin{equation}
\label{master_equation}
U_+-U_-=\alpha\, U_0,\quad \alpha(\omega)
=s\tau_0+\frac1{G_0},\qquad
s\equiv\mathrm{j}\omega,\qquad
f_0=\frac1{2\pi\tau_0},
\end{equation}
where $U_+$ and $U_-$ are the operational amplifier input voltages,
$U_0$ is the output voltage, $G_0$ is a large number, so that frequently $1/G_0$ is negligible, 
$\omega$ is the frequency of the alternating electric field and $\tau_0$ is a small constant with time dimensionality, which is parameterised with the frequency $f_0$ called crossover frequency of the corresponding operational amplifiers.
In our case we have
\begin{equation}
U_1-U_3=\alpha U_4,\qquad
U_5-U_3=\beta U_2.
\end{equation}
For a double operational amplifier $\alpha\approx\beta$.

Problem: Eq.~(\ref{GIC_12345}) is derived assuming that $\alpha=\beta=0.$
For applications in electronics it is necessary the general dependence of the impedance of the converter $Z(Z_1,Z_2,Z_3,Z_4,Z_5;\alpha,\beta)$ to be derived.
Perform the corresponding calculations and check that your function satisfies the condition
\begin{equation}
\lim_{\alpha\rightarrow 0,\;\beta\rightarrow 0}Z(Z_1,Z_2,Z_3,Z_4,Z_5;\alpha,\beta)
=\frac{Z_1Z_3Z_5}{Z_2Z_4}.
\end{equation} 
In short: derive the frequency dependence of the impedance of the general impedance converter using the general equation for the operational amplifiers.

Instruction.
You can use Internet, you can consult professors in physics and electronics anywhere in the world, you can work together.
At the end like a scientific article describe how this result has been obtained, who you are grateful for the consultation, which books or articles you have used or which websites you have visited.
Check whether this problem has not been solved somewhere and give citations of the article that you have remade in your derivation.
Good luck!

\section{Problems for Further Work (Do not read down during the Olympiad)}

\item One of the Olympiad goals is the participants to learn more about the corresponding physics section.
Search literature how different inductances and capacitors are made, what technical solutions there are,
what technology is used, what elements are commercially available to you and think what you are able to do on your own.

\item  The main goal of the Olympiad is to repeat the experiment at home or in the physics classroom in your school.
Ask your physics teacher whether he/she can give you an alternating voltage generator and an AC voltmeter for measurement of small alternating voltages to work with.
We recommend the voltage $U$ applied to the bridge between the points (l) and (r) to be less than 2~Vand the frequency below 20~Hz.

\item Search in Internet in which countries commercial educational equipment illustrating the impedance measurement with bridges is used.

\item If you do not have a voltage generator, use the power supply frequency from 50~Hz (or 60) but in this case you should use a secure transformer.
Turn to your parents.
The voltage which you can use should mandatory be below 36~V; larger voltages are life threatening and are forbidden for usage in high school education!
But for your set-up voltages larger than the power supply voltage of the 9~V batteries are dangerous for the operational amplifiers.
We recommend that voltages larger than 6~V not to be applied to the bridge.
The art of the experimenter is to measure weak signals.
In short, measure again the given inductance.
Compare your results with the measurements of specialized devices for measurements of inductances; look for physics colleagues for assistance.

\item Another goal of the Olympiad is to learn to understand physical and technical articles, see for instance
Refs.~\onlinecite{bridge_story,inductance}.

\begin{figure}[h]
\centering
\includegraphics[scale=0.4]{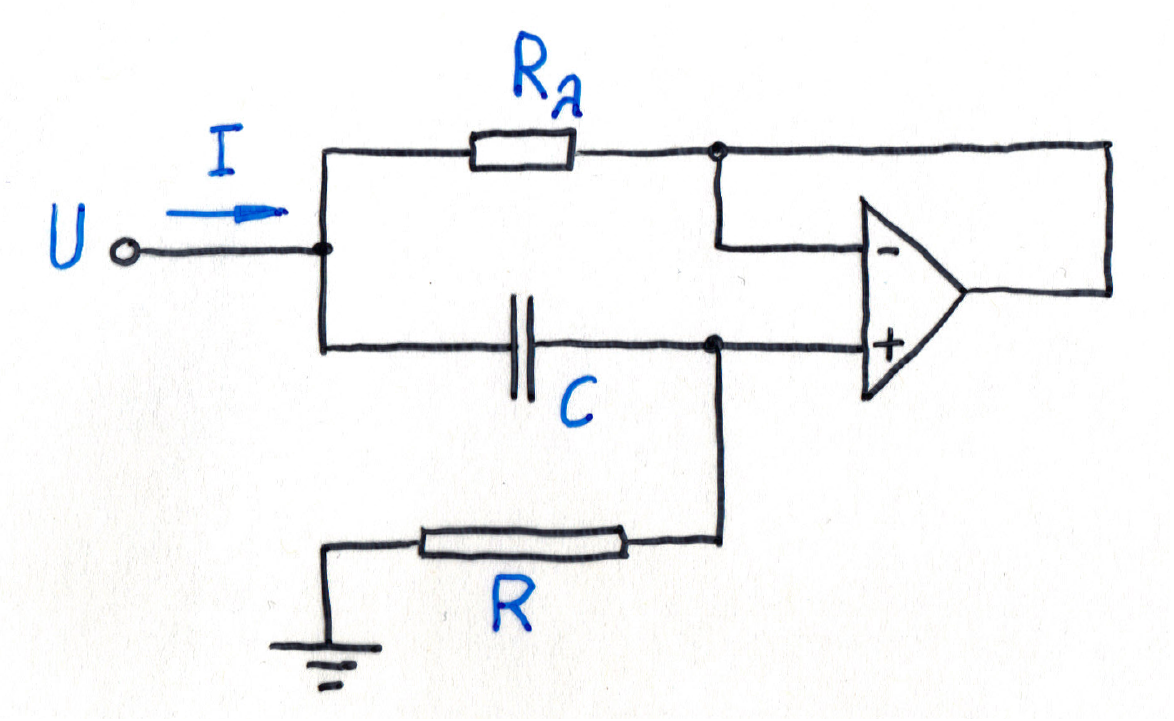}
\caption{The gyrator is the most well-known circuit for simulation of inductance without coils and diamagnetic core and only with two resistors $R$ $R_\lambda$, a capacitor $C$ an an operational amplifier, whose exit current equals the potentials of the voltage inputs (+) and (-), see Eq.~(\ref{girator}). }
\label{Fig:girator}
\end{figure}
The circuit of the so-called gyrator is shown in Fig.~\ref{Fig:girator}.
Use your knowledge that you have acquired at the analysis of the theoretical problems at the Olympiad and derive the approximate formula for the impedance of the gyrator
\be
\label{girator}
Z_\mathrm{g}\equiv\frac{U}{I}
=\frac{R_\lambda+\mathrm{j}\omega R_\lambda RC}{1+\mathrm{j}\omega R_\lambda C}
\approx R_\lambda+\mathrm{j}\omega L_g,\quad
L_g=R_\lambda RC,\quad \omega R_\lambda C\ll1, \quad R_\lambda\ll R,
\quad \frac{L_\lambda}{R_\lambda}=RC. 
\ee
This is another way for making of large inductances but unlike the inductance made by GIC, the inductance of the gyrator has larger ohmic resistance.

\item Think about how the inductance from the Olympiad set-up  can be connected in a high quality resonance contour.
This would be a very good development for the international competition
``Devices for the physics classroom''.\cite{uredi}
We are looking forward to meet you in Sofia on 6 June (Saturday) 2020.

\item Send to the address of the Olympiad your responses, recommendations, criticism and suggestions, which you think would help the EPO7 organizers next year at the same time!

\item When horizontally to the bridge a constant voltage from batteries is applied, only resistors should be used and the voltmeter should be switched to the maximal precision for constant voltage.
Prepare your co-students for a demonstration of a Wheatstone bridge balancing.

\item Turn to your teacher or a colleague physicist and if one has a LCR-meter ask him/her to show you how the inductance and the capacitance from the set-up are measured.
Check the precision of the method in which no specialized devices are used but only a universal multimeter.

\item  The notion inductance is frequently used in the high school courses in physics.
Check in which countries there is a lab exercise in inductance measurement.

\item Take the inductance and the capacitance in your pocket, enter an electronics store as a buyer and measure for two minutes these 2 parameters $L$ and $C$.
What is the precision you achieved at the Olympiad?

\item Invent a technical application of our inductivity 
and implement it.

\item A task for the youngest ones from category (S).
The measurement of the inductance is itself quite simple.
Rotate a potentiometer axis until voltage is minimized.
Ask a senior than you co-student or a teacher to connect the circuit for you and you to make the measurement and the inductance calculation.
Later, when you study what inductance is you will much better understand the new material.

\item At the website of the organizers, the best works are published scanned.
Repeat the experiment carefully and enjoying yourself describe the experiment even better.
The main goal of the Experimental Physics Olympiad is to prepare you for further work.
For a beginning reach the level of the EPO7 champion.
We wish you success.

\end{enumerate}

\section{Epilog}
During EPO7, hundred students tried to measure the inductance of a GIC using a modified Maxwell bridge setup. Of them, more than thirty managed to get some kind of a result, and around twenty of them
measured the inductance withing the expected value.
The absolute champion measured his inductance within one significant digit and solved the theoretical problem of the Olympiad.
This year a new country joined the Olympiad.
Six students from Montenegro took part and received two prizes.
With this the greatest town distance of the Olympiad is from Podgorica, Montenegro to NurSultan in Kazakhstan.

All in all, this Olympiad was a worthy continuation of the now well established tradition that is slowly maturing towards its first big jubilee.

Here the organizers of EPO7 would like to acknowledge everyone who helped in the preparation of this wonderful competition including,
our students who for several days had their lunch and dinner at the faculty: Sofija Jakimovska, Maja Koneva, Verica Mitrevska, Valentina Nikolikj;
The president of the Society of Physicists of Macedonia, prof.~Lambe Barandovski and
Kemet Electronics -- Macedonia for the supply of the operational parts.
We would like also to thank to the friends of EPO -- Peter~Todorov, Martin~Stoev, Nikola~Serafimov, Victor~Danchev and Aleksander Stefanov.

Expecting you all at EPO8.


\appendix

\section{Inductance realized by a general impedance converter}

\begin{figure}[h]
\centering
\includegraphics[scale=0.4]{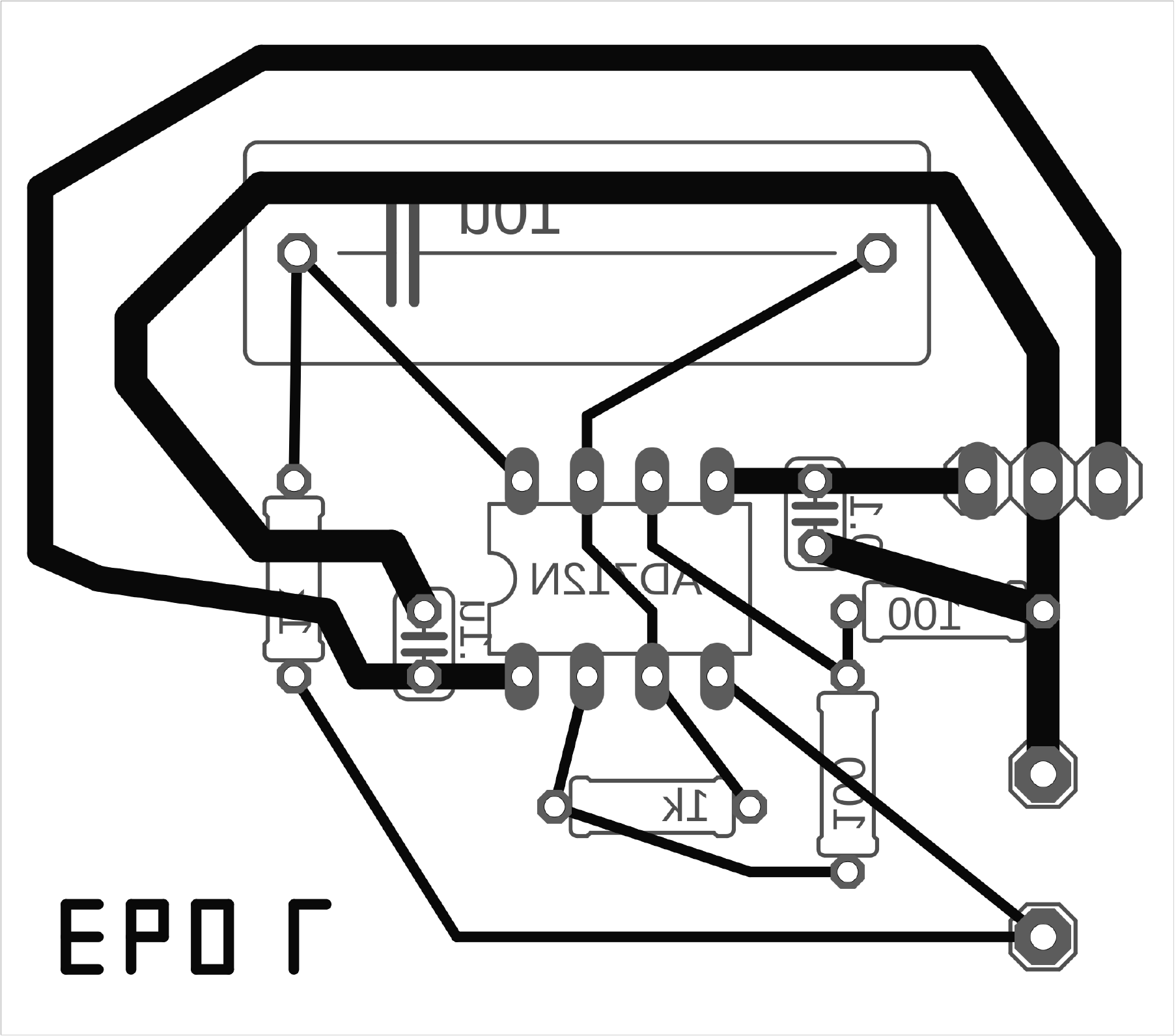}
\caption{A drawing of the printed circuit board through which the inductance of the set-up schematically shown in Fig.~\ref{Fig:GIC} is made.
The wider traces are the power supply and the grounding (the middle point from the 3 ones is with zero potential $U_\mathrm{COM}=0$).
To these 3 points 3 pins are soldered to which a board connector with the conductors coming from the batteries is connected.
The ground is at the same time one of the inputs of the general impedance converter (GIC) through which the inductance from the EPO7 set-up is realized.
The most important GIC input is down to the right where the input voltage $U$ is applied and the input current $I$ enters.
It is easily seen the place of the large capacitor $Z_2=1/\mathrm{j}C.$
The small capacitors are between the ground and the power supply voltages.
The resistor $r_3$ is under the 8 electrodes on which the socket of the operational amplifier 
AD712\cite{AD712} is soldered.
The resistor $r_1$ is placed vertically on the left.
On the right $r_4$ is vertical and $r_5$ is mounted horizontally.}
\label{Fig:board} 
\end{figure}

This text is designated for colleagues designing principally new set-ups for the education.
The following of the traces of the printed circuit board is a good starting lesson in electronics.
The printed circuit board presented in Fig.~\ref{Fig:board} is a typical example how with modest resources the high school education can be significantly renewed.
The cheap highly technological products give the possibility for a significant price decrease of the educational experimental set-ups.
Not a demonstration device in the physics classroom, but set-ups which all children from the class can simultaneously use in a random classroom.

We address the colleagues inventing new problems with a view towards the future.
In the colleges of some countries oscilloscopes are used for a long time and their prices constantly go down.
It is quite realistically in a few years oscilloscopes to be used at the Experimental Physics Olympiad.
For low frequencies the computers themselves can be used as oscilloscopes and this has widely entered the education.
Let us consider what principally new problems can be solved in this process of democratization of the high technology.

\clearpage


\begin{thebibliography}{99}

\bibitem{EPO1}
V.~G.~Yordanov, P.~V.~Peshev, S.~G.~Manolev, T.~M.~Mishonov,
``Charging of capacitors with double switch. The principle of operation of auto-zero and chopper-stabilized DC amplifiers'',
arXiv:1511.04328 [physics.ed-ph]

\bibitem{EPO2}
V.~N.~Gourev, S.~G.~Manolev, V.~G.~Yordanov, T.~M.~Mishonov,
``Measuring Plank constant with colour LEDs and compact disk'',
arXiv:1602.06114 [physics.ed-ph].

\bibitem{EPO3}
S.~G.~Manolev, V.~G.~Yordanov, N.~N.~Tomchev, T.~ M.~Mishonov,
``Volt-Ampere characteristic of "black box" with a negative resistance'',
arXiv:1602.08090 [physics.ed-ph].

\bibitem{EPO4}
V.~G.~Yordanov, V.~N.~Gourev, S.~G.~Manolev, A.~M.~Varonov, T.~M.~Mishonov,
``Measuring the speed of light with electric and magnetic pendulum'',
arXiv:1605.00493 [physics.ed-ph].

\bibitem{EPO5}
T.~M.~Mishonov, E.~G.~Petkov, A.~A.~Stefanov, A~P.~Petkov, I.~M.~Dimitrova,
S.~ G.~Manolev, S.~I.~Ilieva, A.~M.~Varonov,
``Measurement of the Boltzmann constant by Einstein. Problem of the 5-th Experimental Physics Olympiad. Sofia 9 December 2017''
arXiv:1801.00022v4 [physics.ed-ph].

\bibitem{EPO6}
T.~M.~Mishonov, E.~G.~Petkov, A.~A.~Stefanov, 
A.~P.~Petkov, V.~I.~Danchev, Z.~O.~Abdrahim, Z.~ D.~Dimitrov, 
I.~M.~Dimitrova, R.~Popeski-Dimovski, M.~Poposka, 
S.~Nikoli\'c, S.~Miti\'c, R.~Rosenauer, F.~Schwarzfischer, 
V.~N.~Gourev, V.~G.~Yordanov, A.~M.~Varonov
``Measurement of the electron charge $q_e$ using Schottky noise. 
Problem of the 6-th Experimental Physics Olympiad. Sofia 8 December 2018'',
arXiv:1703.05224v2 [physics.ed-ph].

\bibitem{chopper}
E.~A.~Goldberg, L.~Jules, ``Stabilized direct current amplifier'', Patent US2684999A \\
\url{https://patents.google.com/patent/US2684999A/en}.

\bibitem{EJP_Planck} 
D.~S.~Damyanov,  I.~N.~Pavlova, S.~I.~Ilieva, V.~N.~Gourev, 
V.~G.~Yordanov and T.~M.~Mishonov
``Planck's constant measurement by Landauer quantization for student laboratories'',
Eur.~J.~Phys. \textbf{36}, 055047 (2015).

\bibitem{EJP_light} 
T.~M.~Mishonov, A.~M.~Varonov, D.~D.~Maksimovski, S.~G.~Manolev, 
V.~N.~Gourev and V.~G.~Yordanov,
``An undergraduate laboratory experiment for measuring $\varepsilon_0$, $\mu_0$ 
and speed of light c with do-it-yourself catastrophe machines: 
electrostatic and magnetostatic pendula'',
Eur.~J.~Phys. \textbf{38}, 025203 (2017).

\bibitem{EJP_Boltzmann} 
T.~M.~Mishonov, V.~N.~Gourev, I.~M.~Dimitrova, N.~S.~Serafimov, 
A.~A.~Stefanov, E.~G.~Petkov and A.~M.~Varonov
``Determination of the Boltzmann constant by the equipartition theorem for capacitors'',
Eur. J. Phys. \textbf{40}, 035102 (2019).

\bibitem{EJP_Schottky}
T.~M.~Mishonov, E,~G.~Petkov, N.~Zh.~Mihailova, A.~A.~Stefanov, 
I.~M.~Dimitrova, V.~N.~Gourev, N.~S.~Serafimov, V.~I.~Danchev, and A.~M.~Varonov,
``Simple do-it-yourself experimental set-up for electron charge $q_e$ measurement'',
Eur.~J.~Phys. \textbf{39}, 065202 (2018).

\bibitem{TeachSpin}
Teachspin Inc., \url{https://www.teachspin.com/}.

\bibitem{master}
T.~M.~Mishonov, V.~I.~Danchev, E.~G.~Petkov, V.~N.~Gourev, I.~M.~Dimitrova, 
N.~S.~Serafimov, A.~A.~Stefanov and A.~M.~Varonov,
``Master equation for operational amplifiers: stability of negative 
differential converters, crossover frequency and pass-bandwidth'', 
J. Phys. Commun. \textbf{3}, 035004 (2019).

\bibitem{bridge_story}
``A History of Impedance Measurements'' \url{https://www.ietlabs.com/pdf/GenRad_History/HenryHall/HistoryImpedanceMeasurements/PART\%201\%2091507\%20ins\%20pix.pdf}.

\bibitem{inductance}
``Inductor'' at Wikipedia, \url{https://en.wikipedia.org/wiki/Inductor}.

\bibitem{uredi} 
``Devices for the physics classroom'', \\
\url{https://sites.google.com/a/bgphysics.eu/bgphysics/deynosti/konkurs-uredi-za-kabineta-po-fizika}.

\bibitem{AD712}
Analog Devices, ``Precision, Low Cost,High Speed BiFET Dual Op Amp'', Rev.~I \\
\url{https://www.analog.com/media/en/technical-documentation/data-sheets/AD712.pdf}.

\bibitem{Mazda:87}
F.~F.~Mazda, \textit{Electronic Instruments and Measurement Techniques} 
(Cambridge University Press, New York, 1987), Fig.~7.6.

\bibitem{Pramanik}
S.~Pramanik,
``A.C. Bridges for Measurement of Resistance, Inductance, Capacitance, Frequency etc.'',\\
\url{http://www.bengalstudents.com/contents/AC\%20Bridges\%20for\%20Electrical\%20Measurement}.

\bibitem{girator}
G.J.~Deboo, ``Gyrator type circuit'', US patent 3493901, filed 5 March 1968,
issued 1970-02-03, assigned to NASA \\
\url{https://patents.google.com/patent/US3493901}; \url{https://en.wikipedia.org/wiki/Gyrator}.

\bibitem{GIC}
``General impedance converter '', Patent CN86205529U
\url{https://patents.google.com/patent/CN86205529U/en} (1986).

\end{thebibliography}
\end{document}